# India's Rise in Nanoelectronics Research

Udayan Ganguly, Sandip Lashkare, and Swaroop Ganguly, IIT Bombay

## Abstract

Modern semiconductors innovation has a strong relation to scale and skill. While India has a significant demand for semiconductors, it has a daunting challenge to create a semiconductor ecosystem. Yet, India has quietly come a long way. Starting with Centers of Excellence in Nanoelectronics (CENs) initiated in 2006 and broad science and technology funding, India has transformed its nanoelectronics research ecosystem. From negligible contributions as late as 2011, India has risen to be a top contributor to IEEE Electron Devices journals today.

Our study presents important observations in terms of ecosystem development. First, there is a 6-year incubation time from infrastructure initiation to first papers. Then, 4 more years to become globally competitive. Second, growth in experimental research is essential along with modeling & simulations. Finally, the aspirational goals of translational research to contribute to the global technology roadmap requires cutting-edge manufacturing infrastructure & ecosystem access – which still needs development.

The learning informs a call to action for the research ecosystem i.e. academia, industry, and policy-makers. First, sustain and amplify successful strategies of national research infrastructure & funding growth. Second, enhance international collaborations to add further scale & infrastructure to R&D. Finally, strengthen the industry-academia-policy consortium approach to transform to an innovation-based economy. Ultimately, the electron devices community is entering an exciting phase where "Beyond Moore" offers open opportunities in materials, devices to systems, and algorithms. India must build on its success to play a significant role in this new world of disruptive innovation.

## Witnessing a Quiet Evolution

As semiconductors innovations powers the digital age, India has aspired to contribute to nanoelectronics. Towards this, India pushed a strong nanoelectronics program starting in 2006. Our study shows that recently India has been significantly contributing to electron device society related publications. This evolution has been quiet, largely unnoticed in the humdrum of academic and research life. This paper presents the development of the Indian nanoelectronics R&D ecosystem from nothing to global competitiveness. The lessons from this shared experience will help India and other countries who aspire to set sail for similar R&D adventure develop a more grounded, and robust policy and implementation towards an ambitious future.

### A snapshot from the past

In 2004, the industry was pushing transistor scaling to power faster & denser electronics at Intel and AMD. In academia, Cornell and Stanford nanofabrication facilities were driving nanoscience and technology – pushing cutting-edge nanoscale physics from carbon nanotube and quantum dot to the technology of deep-sub-micron CMOS. Indian students would naturally evolve into contributors in international academia to the semiconductor industry with rewarding careers. Yet, many dreamed to pursue a career in nanoelectronics research in India. However, semiconductors research was not at a similar intensity in India. Essentially, engineering institutes in India barely had a microfabrication facility.

At the time of graduation, some students would embark on a pilgrimage to Indian research-centric institutes to get the "lie of the land" – a core instinct for experimentalists. The news was that Indian Institute of Science (IISc) and Indian Institute of Technology (IIT) Bombay had proposed two Nanofabrication Facilities – one at each location. The effort was being led by the Office of Principal Scientific Advisor through the Ministry of Electronics and IT (MeitY). It was a big relief. Yet, there were questions. Will academia be able to develop the Nanofabs? Will new faculty be able to develop a globally competitive R&D program around these new Nanofabs? Only time could tell…

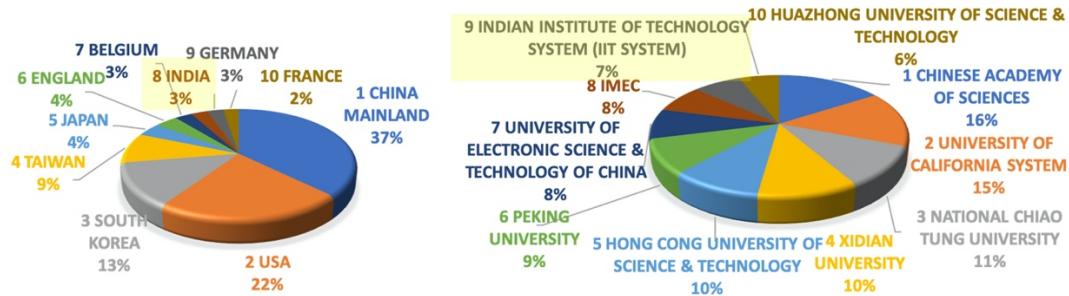

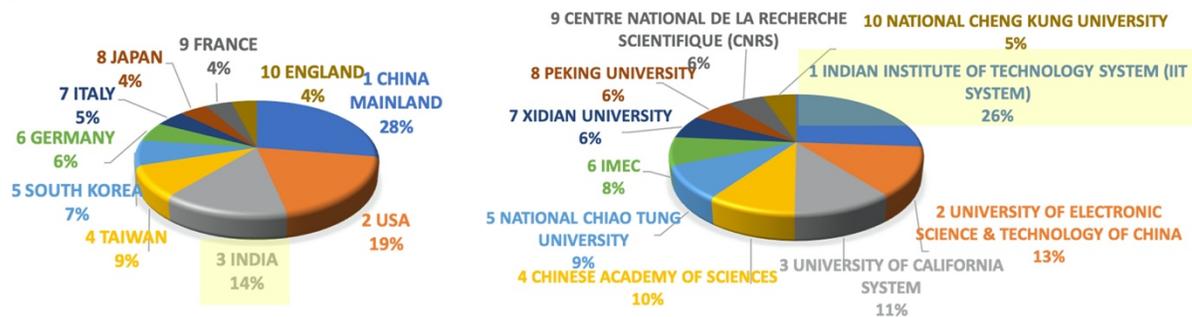

Fig. 1: India and the IIT system records top 10 performances in 2019 including the IIT system being #1 in TED. (Pie chart is limited to the top 10 contributors only)

Recent Indian contributions Electron Device Journals

Fast-forward to 2020, when an IEEE Electron Device Letters (EDL) journal annual performance report analyzed its contributors. In a corner of the report, a strange was a little graph caught the eye. It showed that India is ranked 8th in the contribution by countries and IIT System is ranked 9th. Further IITs have contributed 34/40 papers towards the Indian contribution. This was curious. In the whirlpool of research, funding, and teaching, and fighting administrative fires, most academics would have no bandwidth left to contemplate the "overarching" state of Indian R&D – except in anecdotes. But, this nugget *had* to be investigated.

Among many questions, the dominant one is whether EDL is an anomaly. So, we looked up the IEEE Transactions on Electron Devices (TED) performance. It was a further surprise. IIT system is ranked 1st in the contribution by organizations and India is ranked 3rd. These were compelling indicators. Both EDL and TED are considered the most exclusive venues to publish electron device-related research. India was doing well in both.

## A Trajectory of Strong Growth

The next question was – how long has this been going on? A comparative study of the University of California (UC) System, Chinese Academy of Science (CAS), National Chiao Tung University (NCTU) Taiwan, and the Indian Institute of Technology (IIT) System revealed the trajectory. Unlike these systems (UC, CAS, NCTU) which have been publishing regularly, the IIT System had negligible publications before 2010. Since 2010-2015, the IIT system has *accelerated* publication in IEEE TED as well as in IEEE EDL. The publication in TED has overtaken other university systems and is ranked 1st in 2019. However, in IEEE EDL the publication rate has been comparable. S*aturation* has been observed starting from 2016- present, which has caused the publication rates to flatten – leading to consistent performances in the past 4 years. Further, the citations per paper were comparable to the journal average – slightly better than NCTU and CAS but poorer than UC (Fig. 2).

While India had significant achievements even as early as 2005 in basic science ranking in SCOPUS etc. [1], the growth in nanoelectronics engineering research from negligible contributions before 2011 to high intensity by 2018 is a significant success for the program.

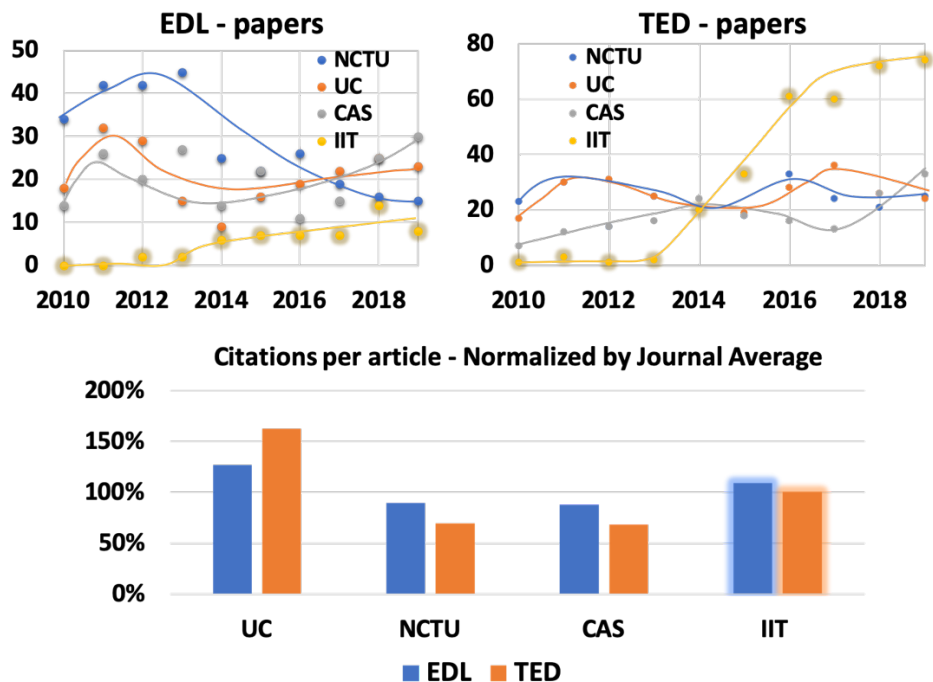

Fig. 2. A timeline of IIT publications in EDL and TED shows negligible publications before 2011. A sharp growth is observed in 2011-2016 followed by a slowdown in growth and saturation 2016 onwards. As a measure of relevance of papers, the IIT system has an average citation per article of about the same as the respective journal averages - better than CAS and NCTU - which are similarly aged institutions but lower than UC System - which is an older player. (Based on data from Web of Science database 2010-2020)

# Resolving Growth

As the growth trajectory is apparent, the next question is - what are the engines of the change? We need to resolve by institutions, collaborations, approach (experiment vs modeling), and individual. Additionally, what is a representative measure of the change? We chose publications and citations per article – with the sense that one could always follow up with a more detailed study. Lastly, what publications do we study to resolve a cogent trend? Here, we chose three publications in the domain of electron devices to study the progression from detailed studies (TED) to newsworthy (EDL) to industrial translation (a conference known as International Electron Device Meeting i.e. IEDM) – where the community showcases the latest and greatest papers of industrial impact. While this list is not exhaustive, we hoped it would throw up representative trends and questions – which may motivate a more specific or exhaustive study.

## Intensifying Experiments to support Modeling & Simulations Expertise

In the microcosm of electron devices, we assessed the contributions of the various Indian institutions by comparing the performance in three publications - EDL, TED, and IEEE International Electron Device Meeting (i.e. IEDM). The papers considered are published between the year 2000-2020.

The foundational publication that is TED, values detailed, long-form presentation on ideas clarifying existing questions. Between 2000-2020, about 473 papers were published in TED of which 131 papers (28%) are experimental. Thus, a vast majority of papers are based on modeling and simulations. This seems to indicate that India contributes strongly to the physics-based modeling of devices in TED.

In comparison, the venue to highlight novelty and relevance, EDL, values concise, newsworthy articles. Between 2000-2020, about 68 papers are published in EDL of which 54 papers (79%) are experimental. Thus, it appears that novel ideas require experimental demonstration to establish newsworthiness. It stands to reason that significantly "surprising" ideas are more credible when presented experimentally.

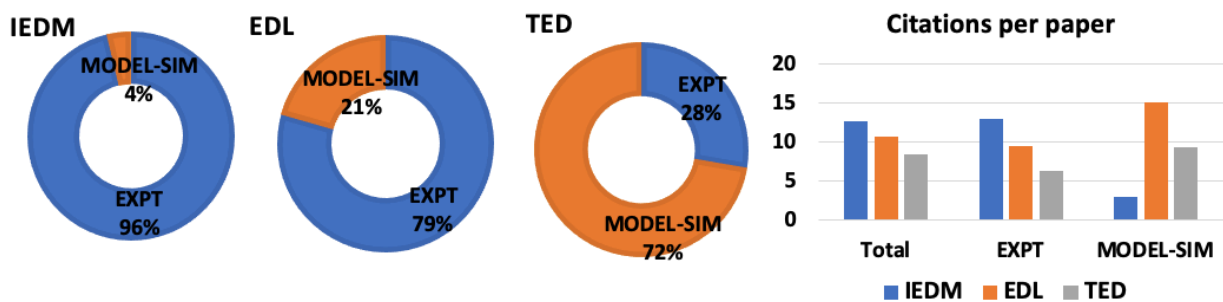

Fig 3: Papers published between 2000-2020 show that the experimental focus is increasingly essential to acceptance going from TED to EDL to IEDM – which is in the order of increasing citations per paper.

Finally, as a venue to inspire a community that has created the electronics revolution through the relentless miniaturization in size and cost – whose essence is Moore's Law, IEDM is the flagship conference where industry & academia showcase the latest and greatest technological development of industrial relevance. As much as 96% of the papers are based on experimental demonstrations (Fig. 3). The average citations per paper are about similar (approx. 10.5) for both IEDM and EDL and 8.43 for TED. *The strong requirement for experiments to breach the next levels of impact is clear as day.*

## Institutions of Success

The IITs and IISc have the reputation of being top engineering education institutes [2]. IISc has been a postgraduate centric institute while IITs were undergraduate centric to start with but recently evolving towards post-graduate research focus. The IITs are a heterogeneous lot. The old IITs (Kharagpur, Bombay Madras, Delhi, and Kanpur) established in the 1950s have been transformative for India as well as the international technology ecosystem. To replicate the success, various new IITs were created and some older engineering schools like those at Roorkee and Benares Hindu University (Varanasi), were brought into the IIT system.

Among these top institutions, IIT Bombay and IISc Bangalore are the top contributors. Coincidentally, they also have the oldest Center of Excellence in Nanoelectronics i.e. the state-of-the-art experimental facilities. About 13 IITs and IISc published 68 papers in EDL while about 17 IITs and IISc published 473 papers in TED between 2000-2020. This smaller number of publications in EDL by few IITs can be directly related to the state-of-the-art experimental facilities available in the top IITs and IISc. In comparison, 4 IITs and IISc published 26 IEDM papers – *a gap to aggressively improve upon.* (Fig. 4)

While, the top contributors are primarily the older IITs & IISc, newer IITs like Indore, Guwahati, Jodhpur, and Gandhinagar to name a few have started contributing strongly. Further, older institutions but newer entrants to the IIT system like Roorkee & Varanasi have also been quite productive. This overall performance has been encouraging.

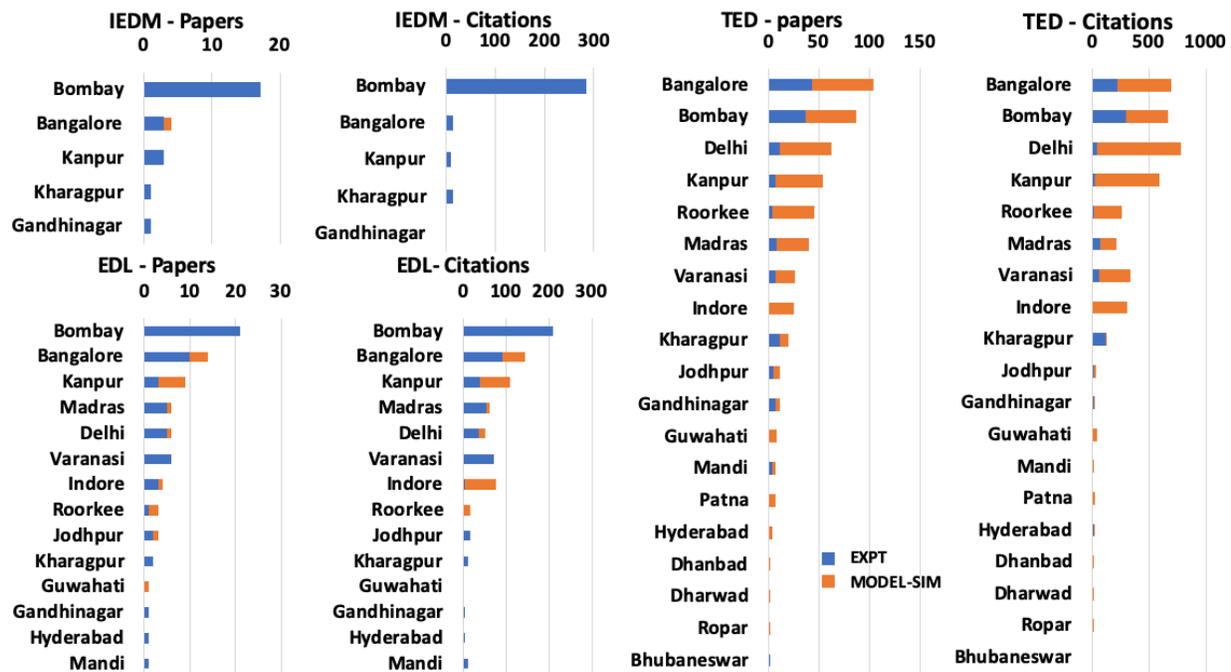

Fig. 4: IEDM is the most selective with 5 institutes boasting of 26 papers while EDL and TED have 14 and 18 institutes publishing 68 and 473 papers respectively between the year 2000-2020. IIT Bombay and IISc Bangalore top the list.

## Independence and Collaboration

In academic research, every new assistant professor dreams to set up a lab that propels "ideas" that become major contributions to capture the imagination of the community of researchers and technologists. The *ultimate* reward is that the ideas may translate from research to manufacturing. At the early stages, the ideas are typically driven by a single or few groups within the country. As the idea grows, collaborations become necessary with industry or international academia. The publication data was resolved into three parts – (i) Solely Indian, (ii) Indian with International University collaboration, and (iii) Indian with industry collaboration. First, *solely Indian* papers correspond to the papers in which all the authors are affiliated with Indian institutes – which indicates that the talent and infrastructure required is available within the country. Second, "*Indian with International University*" corresponds to the papers by authors affiliated with Indian institutes collaborating with international universities. Thirdly, *Indian with industry* corresponds to the papers where authors are affiliated with Indian institutes have collaborated with the industry anywhere in the world. Categories (ii) and (iii) show the complementary skills of Indian researchers required by international academia and industries.

"Solely Indian" papers are 2.7x and 4x higher than "Indian with international university collaborations" for EDL and TED respectively. Slightly lower fraction of Indian papers with Industry is observed. In contrast, most of the papers are either Indian with international university collaboration or Indian with industry for IEDM. The citations per paper follow the descending order of IEDM to EDL to TED - except solely Indian papers at IEDM – which has a slightly lower average citation per paper. This may be correlated to more recent papers getting lesser citations (Fig. 5).

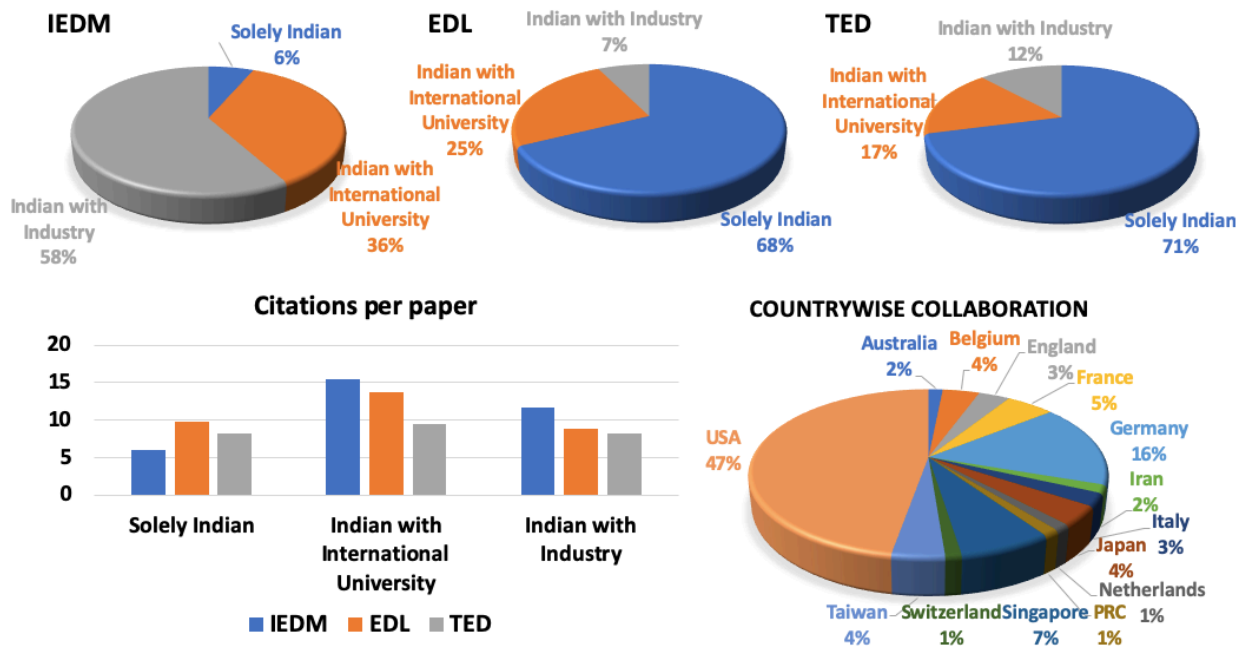

Fig. 5: (a) "Solely Indian" papers are compared with collaborations of Indians with International Universities and Industries. It highlights the availability of independent national talent and infrastructure for EDL and TED. However, collaborations are largely essential for IEDM. Similarly, the citations per paper of "solely Indian" papers quite at par with "Indian with Industry". The collaboration of Indian institutes with researchers from other countries shows strong international ties.

Among the international affiliation of collaborators, the top 3 countries are the USA, Germany, and Singapore. This also indicates opportunities to strengthen ties with countries like Belgium, Taiwan, Brazil, Japan, and South Korea.

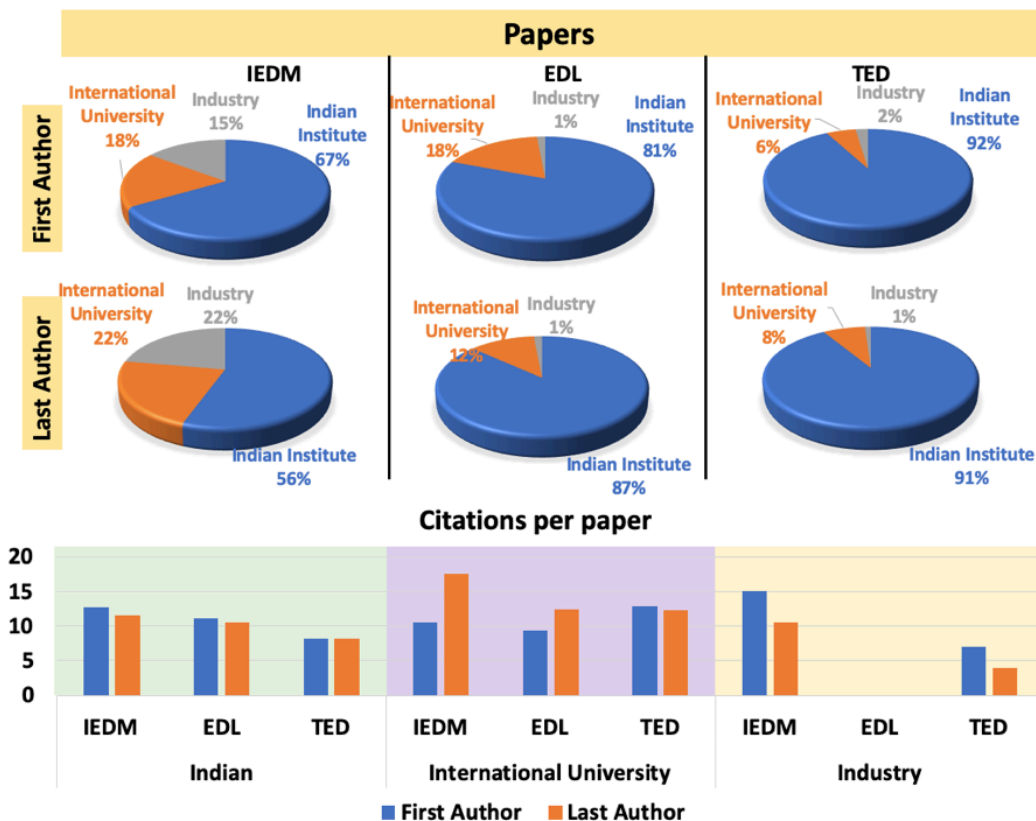

Fig. 7 (a) The comparison between affiliations of first and last authors papers with (b) citations per paper.

### The frontline talent (first author) vs. the science manager (last author)

A talented scientist starts as a graduate student, poring through details like pushing ideas, and their execution crystallizing into impactful papers. Such a scientist ultimately transforms into the butterfly of a science manager – the professor or research lead i.e. the last author, who drives research programs with many such talented scientists. "Indian Institute" contributions resolved by first authors vs. last authors indicate major Indian contributions – as opposed to non-first or non-last authors, who are contributors but not leads. Similarly, "International University" and "Industry" contributors are also analyzed for first vs. last authors.

When comparing only the first authors, typically the fraction of Indian first authors are high for TED and EDL (92% & 81% respectively) but falls sharply for IEDM to 67%. The international academic first authors increase from 6% & 18% for TED and EDL respectively to 18% for IEDM. Industrial first authors are negligible for TED and EDL but increase to 15% for IEDM. The role (or even necessity) of industry collaboration for IEDM is again highlighted – unlike for EDL or TED.

The last author statistics are similar to the first author for TED and EDL. However, 56% of Indian last authors compared to 67% of Indian first authors indicating that a fraction of the IEDM papers associated with India is "managed" by international academia or industry (Fig. 7). This further highlights the reliance on *industry-relevant* experimental capabilities and vision – presently a challenge in India. This may be due to the weakness of local semiconductor manufacturing R&D, which is mitigated by the strong connections of Indian researchers with the international community.

The average citation per paper again follows the IEDM to EDL to TED in descending order – with a very small difference between first and last authors of a specific type. However, international academic last authors at IEDM have significantly higher citations per paper than their counterpart among first authors – while the relationship is flipped for industrial first vs last authors in TED. Overall, first and last authors from Indian institutes show competitive citations per paper and significant contributions in TED and EDL with IEDM as the next goal – and hence the focus of community strategy.

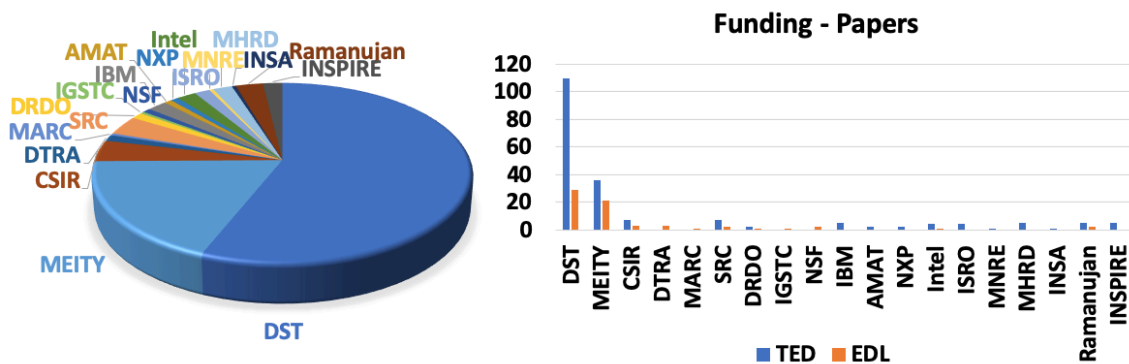

Fig. 8 Papers affiliated with funding agencies and the resolution of papers published in EDL vs. TED is shown.

## Policy Drivers of Growth

With the growth trajectory and engines resolved, the drivers of growth require enumeration, as they identify the strategies to further support and grow.

### Increase in Institutes, Faculty and Graduate Student Hiring

The engineering education infrastructure produces a vast number of engineers [3] – who compete for a very limited number of top graduate research positions [4]. This provides an excellent pool of talent for Indian research with aggressive competition. The growth of the academic ecosystem across the country to harness this strong supply of talent is a critical reason. As an indicator, we looked at IIT Bombay hiring in Electrical Engineering broadly from 2006 to 2013 as a correlated measure. Here, the faculty strength grew from 37 to 58 i.e. almost a 60% increase in the 8 years. At the same time, Ph.D. students in IIT Bombay grew from 116 to 422 from 2007-2019 which is almost 4x growth. The number of IITs grew from 8 to 23 from 2008-2019 which is a 3x growth. These are excellent trends in higher education infrastructure [5].

### Consolidated Nano-Centers funding create key experimental Infrastructure

In 2006, the Center of Excellence in Nanoelectronics (CENs) program was set up by the Ministry of Electronics and Information Technology (MEITY) championed by the Office of Principal Scientific Advisor to the Government of India. IIT Bombay & IISc Bangalore were the first institutes to get a Nanofab in 2006. The model was later expanded to new centers in Madras, Delhi, and Kharagpur. Kanpur has developed an equivalent thrust in displays. From 2018 onwards, the Department of Science and Technology (DST) joined hands with MEITY to co-fund the CENs. The Indian Nanoelectronics Users program was initiated in 2009 to support the usage of the CENs by nanoelectronics researchers across the country.

### A Scatter of Project-based Funding creates Talent Pool and Innovation

Both MeitY and DST continued and even enhanced funds for smaller projects for researchers across the country to create a talent pool. These smaller labs have created the development of diverse ideas – critically essential for academic nimbleness – at the efforts requiring smaller lab infrastructure e.g. in aspects of materials, physics, chemistry science, - especially in simulations & modeling. The most promising ones could be funneled into the integrated laboratory platforms like CENs for further evaluation or application into integrated devices.

### Correlation of Papers to Funding Source

An analysis of funding sources of papers shows that DST is the highest funding affiliate of paper from India. However, resolving papers by journal shows that MEITY funding has produced as many EDLs as DST funding, while DST funding produces a large fraction of TED papers related to modeling/simulations (Fig. 8).

### Relatively Modest Funding converted to High-Efficiency Research

Indian funding started in 2006 with the Center for Excellence in Nanoelectronics with funding till 2020 is approximately 100 Million USD by MeitY. US started the National Nanotechnology Infrastructure Network (NNIN) program in 2000 and the funding till 2016 was about 22 Billion USD [6] - a factor of 200x more than India. Further, India spends overall 0.7% of its GDP in R&D compared to 2% average and 4.5% for Korea or Israel i.e. 3-6x lower [1]. The correlation to the productivity of the researchers in India simply comes from the excellent talent in the ecosystem. This has also forced India to develop & succeed in modeling & simulations as opposed to experiments – where infrastructure is needed. There is no surprise here. However, while research can be in a silo to an extent, technology translation requires a stronger network – which requires a quantum jump in the funding, infrastructure, and policy to incentivize behavioral changes in the ecosystem. We can look at the history of CENs to review a successful experiment in the research ecosystem creation in India.

### Incubation & Development Time

It appears that the CEN Phase 1 from 2006-2011 produced key infrastructure without resulting in publications. This was followed by a period of productivity in Phase 2 (2011-2016) – which coincided with the maturing of the research ecosystem. The upswing in publications started around 2012 which indicates a *6-year incubation timescale*. Such a timescale may be resolved informally into 3 years of lab creation, 2 years of faculty hiring on the announcement of a working nano-center followed by 2 years to get training. Ultimately it took till 2016 to show productivity – which is a 10-year development period to global competitiveness.

### Shared Nano Center enable effective / efficient R&D utilization & productivity

Further, the strong "experimental" EDL papers in IIT Bombay as an indicator may be correlated to DST-funded equipment placed in the shared MEITY funded CEN facilities where the integrated facilities were more effectively maintained and shared in producing experimental EDL papers. Equivalent correlations may be observed in other institutes.

### National & International Research & Industrial Network Built

While there is a significant volume of R&D done nationally through the formal INUP program, there is also a sense of community with international collaborations. Leading international collaborations are with Berkeley, Ohio State, Albany in the USA, Stuttgart in Germany. Further, various companies (Intel, IBM, NxP, Applied Materials) and consortia like SRC have engaged with researchers at Indian institutes.

### Great Opportunity: Transitioning from More-Moore to Beyond Moore

Moore's Law and scaling has been excellent for the industry but has essentially left the academia to contemplate *sustaining innovation* – which is all about improving existing approaches. With the slowdown of Moore's Law, *disruptive innovations* have taken center focus with More-than-Moore with 3D integration, sensors, 5G RF devices. However, the big opportunity is with the "Beyond Moore" ideas of Artificial Intelligence and Neuromorphic Computing that needs significant and rapid path-finding in the entire value chain of materials-devices, circuits-systems, and algorithms. This provides a rich opportunity for competent research groups, companies, and countries to define the future trajectory in a large unexplored vista of great promise. Multiple inflections in technology, driven by disruptive innovations out of the knowledge ecosystem are expected. In such a world, university R&D is reclaiming its center stage and India must aspire to play a significant role.

## Learnings & A Call to Action

The Indian story is that of remarkable success. It has taken 10 years to catch up with the top academic systems in terms of publications in flagship IEEE journals. So, is it time to relax and cool one's heels? On the contrary – what is built can whither under neglect or bloom with care. So, what should India do? Below, we highlight three thrusts to take nanoelectronics R&D to the next level.

### 1. National R&D Ecosystem is fulfilling its promise; Sustain & Amplify

The Indian research ecosystem has made significant strides. The basic funding model i.e. large "shared" infrastructure funding and support has worked and needs to be sustained. We must remember that *scale & skill are the keys*. This is just like the exponential growth described in Moore's Law sustained by the scale of investment, infrastructure & ingenuity. Similarly, R&D in a little lab is essential for disruptive innovation. But it does not translate well in a vacuum and needs the larger infrastructure to bloom. In the Indian context, it took consolidated shared & scaled up infrastructures (e.g. CENs) to kick-start the semiconductors research ecosystem – which needs to be sustained and amplified.

### 2. Semiconductor Manufacturing Related R&D has green shoot; Grow

Semiconductor manufacturing is complex. Yet, a single step innovation in a 200+ step manufacturing process may be the next disruptive innovation. It may create a cascade of opportunities at all levels from design to hardware. Yet, it cannot even be tested without investing in the baseline of 200 steps. While initial feasibility can be shown in CENs, a translation center is required. IIT Bombay has a small AMAT nanomanufacturing lab processing 8" wafers – which has developed cutting edge germanium transistor technology. To take this to the next level, an R&D center (like IMEC Belgium, ITRI, Taiwan, IME, Singapore), which are essential to translate existing research to industrially relevant technologies. We must remember that it took 6 years to incubate the research ecosystem and 15 years to reach significant productivity. An R&D center may require a 15-20-year vision.

### 3. International & Industrial Networks are formed, Strengthen

We observe significant interactions with various international academics and industries. We need to take these to the next level with bi-lateral collaborations and funding. Further, industry-academia joint funding models where SRC collaborates with the Government of India to develop co-funding and R&D strategy will be pertinent for technological relevance.

## Conclusions

A beautiful bloom has unfolded before our eyes in the past 15 years. It correlates to key policy and implementation of shared Centers of Excellence in Nanoelectronics as rallying points for research talent which was grown all over the country due to a scatter of project funding. A 6-year incubation period followed by an 8-year growth period is observed. Such a successful vision, policy, and implementation should serve to inspire the stakeholders in Indian semiconductors R&D to replicate, refine and scale the model to respond to the urgent national and international needs of the innovation-led semiconductors ecosystem in India.

## Acknowledgments

The study was inspired by data shared and by Prof Jesus del Alamo, MIT, and the Editor-in-Chief for EDL followed by various discussions. We acknowledge discussions with Sunita Verma, Sangeeta Semwal, and Nishit Gupta from MeitY, and Milind Kulkarni from DST. Further, we appreciate the various feedback from Prof. Juzer Vasi, IIT Bombay.

## References

[1]. Department of Science and Technology, Ministry of Science & Technology, Government of India, "Research & Statistics at a Glance", 2020
[2]. "QS India University Rankings", QS Top Universities, 2019.
[3]. Neeti Nigam, "The rise and fall of the Indian Engineering Degree", Indian Express, 2020
[4]. Shyna Kalra, "Over 15% MTech Seats go vacant at IITs as GATE-based recruitment, admissions fail to synchronize", Indian Express, 2020
[5]. "Indian Institute of Technology", Wikipedia
[6]. "National Nanotechnology Initiative: Supplement to the President's 2016 Budget", nano.gov, 2016